\documentclass[a4paper,twocolumn,11pt]{quantumarticle}
\pdfoutput=1
\usepackage[utf8]{inputenc}
\usepackage[english]{babel}
\usepackage[T1]{fontenc}
\usepackage{tikz}

\usepackage{graphicx}
\usepackage{color}
\usepackage{braket}
\usepackage{verbatim}
\usepackage{amsmath}
\usepackage{amssymb}
\usepackage{amsfonts}
\usepackage{mathtools}
\usepackage{braket}
\usepackage{float}
\usepackage{hyperref}
\usepackage{lipsum}
\usepackage{dsfont}
\usepackage[ruled,vlined,linesnumbered]{algorithm2e}
\definecolor{darkblue}{RGB}{0,0,170}

\SetCommentSty{mycommfont}
\DontPrintSemicolon

\begin{document}

\title{Breadth-first graph traversal union-find decoder}
\author{Matthias C. L\"obl}
\email{matthias.loebl@nbi.ku.dk}
\affiliation{Center for Hybrid Quantum Networks (Hy-Q), The Niels Bohr Institute, University of Copenhagen, Blegdamsvej 17, DK-2100 Copenhagen {\O}, Denmark}
\author{Susan X. Chen}
\affiliation{Center for Hybrid Quantum Networks (Hy-Q), The Niels Bohr Institute, University of Copenhagen, Blegdamsvej 17, DK-2100 Copenhagen {\O}, Denmark}
\affiliation{Quantum Engineering Centre for Doctoral Training, University of Bristol, Bristol, United Kingdom}
\author{Stefano Paesani}
\affiliation{Center for Hybrid Quantum Networks (Hy-Q), The Niels Bohr Institute, University of Copenhagen, Blegdamsvej 17, DK-2100 Copenhagen {\O}, Denmark}
\affiliation{NNF Quantum Computing Programme, Niels Bohr Institute, University of Copenhagen, Blegdamsvej 17, DK-2100 Copenhagen {\O}, Denmark.}
\author{Anders S. S\o{}rensen}
\affiliation{Center for Hybrid Quantum Networks (Hy-Q), The Niels Bohr Institute, University of Copenhagen, Blegdamsvej 17, DK-2100 Copenhagen {\O}, Denmark}

\begin{abstract}
Fast decoding algorithms are decisive for real-time quantum error correction and for analyzing properties of error correction codes. Here, we develop variants of the union-find decoder that simplify its implementation and provide potential decoding speed advantages. Furthermore, we show how these methods can be adapted to decode non-topological quantum low-density-parity-check (qLDPC) codes. All the developed decoders can directly include both qubit erasures and Pauli errors in the decoding step, thus addressing the dominant noise mechanisms for photonic quantum computing. We investigate the strengths and weaknesses of the different decoder variants, benchmark their speed and threshold error rates on several codes, and provide the used source code.
\end{abstract}

\maketitle

\section{Introduction}
Correcting errors is crucial for large-scale quantum computing and therefore various error-correction codes~\cite{Dennis2002, Breuckmann2021, Brown2016, Bravyi2024} have been developed and implemented~\cite{Postler2022, Krinner2022, Google2023, Bluvstein2024}. Indispensable for quantum error correction is an accurate and fast decoding algorithm that predicts corrections given error syndromes. Below a decoder- and code-specific error threshold, a sufficiently large code can yield an arbitrarily low logical error rate. Ideally, the decoder achieves a threshold that is as high as possible (high accuracy) such that higher experimental error rates can be tolerated. For this purpose, several algorithms have been investigated~\cite{Brown2023, Roffe2020, Meinerz2022, Higgott2023, Bravyi2014}. The need to reduce logical errors by increasing the code size poses a challenge to such algorithms: the time complexity of the decoding increases with code size, yet the decoding and subsequent quantum error correction needs to happen in real-time~\cite{Terhal2015, Skoric2023, Barber2023}.

Typically, there is a tradeoff between decoding speed and accuracy: a fast and simple decoding algorithm is less accurate, whereas a more accurate decoding algorithm is slower~\cite{Delfosse2023}. Union-find decoders are particularly promising as they yield good threshold error rates and achieve a time complexity scaling linearly with the system size~\cite{Delfosse2021}. At the same time, the low memory requirement of the union-find decoder has allowed the development of architectures running this algorithm on field programmable gate arrays (FPGAs) or application-specific integrated circuits (ASICs)~\cite{Das2022, Liyanage2023, Barber2023, Liyanage2024}.

Here, we investigate several simplifications of the original union-find algorithm. We keep the standard union-find data structure, yet there are two main changes: first, we grow erasure clusters node-by-node using breath-first traversal of the so-called \textit{Tanner graph} of the code instead of growing cluster-by-cluster~\cite{Delfosse2021}. Second, we do not track the boundaries of clusters. We show that identical thresholds can be achieved despite these simplifications, making our decoder potentially faster without sacrificing accuracy. The developed modifications of the union-find decoder can be applied to so-called topological codes~\cite{Kitaev2003, Dennis2002}. Furthermore, we adapt our best decoder variant for topological codes such that it can be applied to non-topological quantum low-density-parity-check (qLDPC) codes. The modifications come at the cost of increased running time when error rates are high, yet fast decoding is maintained when error rates are below the threshold. Importantly, both Pauli errors and erasures are included in the decoding for our qLDPC decoder and for the other decoder variants. Handling erasures in the decoding is a particular strength of union-find decoders for topological codes~\cite{Delfosse2020, Delfosse2021} as it is decisive to compensate for photon losses in measurement- or fusion-based photonic quantum computing architectures~\cite{Raussendorf2001, Bartolucci2021}. Finally, we benchmark our decoders on toric codes as well as bivariate bicycle codes~\cite{Bravyi2024, Panteleev2021, Kovalev2013}, simulating both Pauli errors and erasures. A preliminary version of our decoder is provided as an open-source repository that can readily be used for fast offline decoding of codes with periodic boundary conditions~\cite{Lobl2024}.

\section{Background Information}
All the codes that we consider are in the class of Calderbank–Shor Steane (CSS) codes~\cite{Calderbank1996, Steane1996} which have stabilizer generators $S_G$ and logical operators $\mathcal{L}$ being either products of Pauli $Z$ or Pauli $X$ operators\footnote{The stabilizers $S$ of a code space $\mathcal{C}$ are the cumulative sub-group of the Pauli group such that for all $\ket{c}\in\mathcal{C}$ and for all $s\in S$ the following equation holds: $s\ket{c}=+1\ket{c}$. The stabilizer generators $S_G$ are a generating subset of $S$ with the minimum number of $rank(S)$ elements.}. These stabilizer generators constitute parity checks that can be employed to detect errors. The parity checks of CSS codes can be expressed by two \textit{parity check matrices} $H_X$ and $H_Z$ over $\mathbb{F}_2$ (the binary field with two elements: $0, 1$) and all equations involving these matrices must be interpreted by modulo $2$ in the following. Every row of such a matrix corresponds to one parity check, being non-zero for all the columns corresponding to a qubit of this parity check. $H_X$ represents the parity checks $\mathcal{P}_X$ that are products of Pauli $X$-operators and $H_Z$ represents the parity checks $\mathcal{P}_Z$ that are products of Pauli-$Z$ operators, where $H_XH_Z^T=0$ needs to be fulfilled meaning that all the parity-check operators commute. An equivalent representation of the parity check matrices is the \textit{Tanner graph} which is an undirected, bipartite graph where data qubits, $\mathcal{Q}$, are connected to all parity checks they are involved in~\cite{Breuckmann2021}. Since bit- and phase-flip errors can be decoded independently for CSS codes, we will always consider the sub-graph of the Tanner graph corresponding to either the $\mathcal{P}_X$ or the $\mathcal{P}_Z$ parity checks. When we talk about Tanner graph traversal, we thus mean traversing the corresponding sub-graph of the Tanner graph. We will mostly consider the Tanner graph representation as it provides a natural intuition for union-find decoding. A \textit{cluster} is referred to as a connected component of the Tanner graph and the \textit{distance} between clusters (not to confuse with the code distance) refers to the shortest path between clusters on the Tanner graph.

In the absence of errors, measuring a stabilizer will leave the code space unperturbed and give a value of $+1$. Detectable errors cause some parity checks to be $-1$ and we call these parity checks the \textit{non-zero syndromes}. Measuring stabilizers that are products of Pauli $X$-operators detects phase flip errors and measuring stabilizers that are products of Pauli $Z$-operators detects bit flip errors.

Error correction can be achieved by combining the information from several parity checks to find an error pattern that is compatible with the non-zero syndromes. Let $\sigma_X$ ($\sigma_Z$) represent an $\mathbb{F}_2$ vector with elements being the result of parity checks from $\mathcal{P}_X$ ($\mathcal{P}_Z$). Zero elements represent parity checks with result $+1$, non-zero elements represent parity checks with result $-1$ (non-zero syndromes) indicating the presence of an error. Furthermore, let the $\mathbb{F}_2$ vector $e_Z$ ($e_X$) represent errors on data qubits $\mathcal{Q}$ with non-zero elements corresponding to phase-flip (bit-flip) errors. For a bit-flip error $e_X$, the corresponding syndromes are given by $\sigma_X=H_Ze_X$. For a phase-flip error $e_Z$, the corresponding syndromes are given by $\sigma_Z=H_Xe_Z$. Error correction needs to succeed with only information about which syndromes are non-zero, i.e. using the information in $\sigma_X$ and $\sigma_Z$. The first goal of a decoder is finding correction vectors $c_X$ and $c_Z$ such that $H_Z(c_X+e_X)=0$ and $H_X(c_Z+e_Z)=0$ meaning that the applied operators $(c_X+e_X)$ and $(c_Z+e_Z)$ are within the code space (stabilizers or logical operators of the code). The second and main goal is ensuring that $(c_X+e_X)$ and $(c_Z+e_Z)$ are stabilizers of the code, thus avoiding an undesired logical operator being applied. For a sufficiently large code and below the threshold error rate, this goal can be achieved with an arbitrarily high probability. The threshold error rate depends on the code as well as the employed decoding algorithm.

In the following, we will first consider union-find decoders that apply to so-called \textit{topological codes}, a subset of CSS codes. For such codes, the Tanner graph can be embedded on a manifold such that parity checks are local operations corresponding to a cellulation of the manifold~\cite{Kitaev2003}. Furthermore, every data qubit is shared between exactly two parity checks, meaning that every column in $H_X$ and $H_Z$ has support on exactly two elements. Finally, we will develop a union-find decoder that can be applied to a broader class of qLDPC codes~\cite{Breuckmann2021, Bravyi2024}.

\section{Decoding Algorithms}
The union-find decoder from Ref.~\cite{Delfosse2021} works in two steps: (1) Syndrome validation: growing clusters until they merge and form a valid (decodable) cluster. For topological codes, a cluster is decodable once it contains an even number of non-zero syndromes or has support on the boundary of the code~\cite{Delfosse2021}. (2) Decoding the valid clusters individually with the peeling-forest algorithm from Ref.~\cite{Delfosse2020}. When clusters are local and so do not percolate the code, this decoding algorithm likely does not produce a logical error. Logical errors become likely once a cluster either percolates the code or at least supports a large fraction of a logical operator\footnote{The distance between clusters can be used to estimate the probability of a logical error, which can be a useful soft output of a union-find decoder~\cite{Meister2024}.}.

Below, we will make various modifications to the union-find decoder to allow faster decoding. Our initial modifications only affect the syndrome validation in the first step where we assume topological codes with periodic boundary conditions\footnote{The decoder can be easily adapted to codes with boundaries by declaring a cluster touching the boundary as valid~\cite{Delfosse2021} and starting the construction of the forest data structure in the peeling-forest decoder from the boundary~\cite{Delfosse2020}.}. Instead of storing boundaries of clusters and growing cluster-by-cluster, we perform breadth-first graph traversal on the Tanner graph starting from the non-trivial syndromes, resulting in qubit-by-qubit cluster growth. Secondly, we extend the union-find decoder to decode general qLDPC codes.
For this decoder, we employ a different method to decode clusters since they cannot be decoded with the peeling-forest decoder from Ref.~\cite{Delfosse2020}.

\subsection{Noise model}
\label{sec_noise}
\begin{figure*}[!t]
\includegraphics[width=1.0\textwidth]{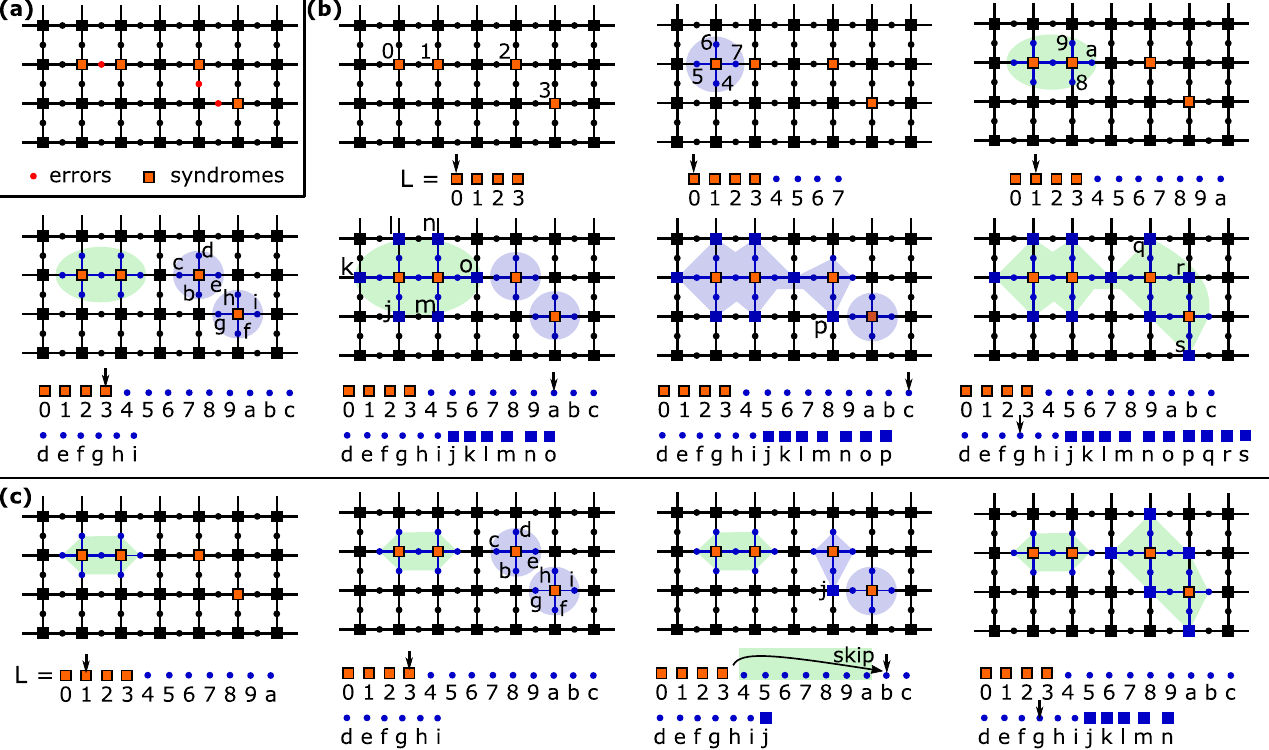}
\caption{\label{fig_intro} Syndrome validation by growing clusters with breadth-first Tanner graph traversal. \textbf{(a)} A patch of the two-dimensional surface code with errors on data qubits shown as red circles and non-zero syndromes as orange squares. \textbf{(b)} Illustration of Algorithm~\ref{alg:1}. All non-zero syndromes constitute a single-node cluster from where cluster growth for syndrome validation is started. For clusters with more than one node, we illustrate valid clusters (with an even number of non-zero syndromes) in green and invalid clusters (with an odd number of non-zero syndromes) in blue. For syndrome validation, all non-zero syndromes are initially added to a list $L$ for the breadth-first traversal of the Tanner graph. The breadth-first graph traversal starts at $L[0]$ (black arrow). For the current element in $L[i]$, the clusters of all its neighbors are merged with the cluster to which $L[i]$ belongs. Furthermore, the neighbors of $L[i]$ are added to $L$ unless they have been added already. Breadth-first graph traversal continues by increasing the position $i$ in the list $L$, going from $L[i]$ to $L[i+1]$. The algorithm terminates once all clusters are valid, having an even number of non-zero syndromes. \textbf{(c)} The main idea of Algorithm~\ref{alg:3}. To avoid growing valid clusters, nodes are skipped in the breadth-first graph traversal if they belong to a valid cluster. This leads to smaller clusters after syndrome validation and makes logical errors less likely.}
\end{figure*}
To demonstrate our decoding algorithms, we consider a simplified noise model where independent and identically distributed (i.i.d.) Pauli errors appear with a uniform \textit{physical error rate}, $p$, and syndrome measurements are perfect~\cite{Delfosse2021}. As bit- and phase-flip errors can be decoded independently for CSS codes, we can assume without loss of generality that $p$ refers to the phase-flip error rate. Additionally, we consider a heralded erasure model where i.i.d. qubit losses of data qubits can happen at known locations with a uniform erasure probability $\epsilon$. Ancilla qubits for syndrome extraction are considered to be unaffected by erasures. Since loss corresponds to a measurement without knowing the measurement outcome, we assign the erased data qubits a random value by applying a Pauli error with a probability of $0.5$~\cite{Delfosse2021}. This model is motivated by photonic quantum hardware, where the loss of photonic qubits is a common imperfection that is detected when measuring the photons. In fault-tolerant measurement-based~\cite{Raussendorf2006} or fusion-based~\cite{Bartolucci2021} photonic quantum computing architectures, all qubits are sequentially measured and syndromes are calculated from the measurement results. Heralded qubit erasures thus occur naturally, whereas there are no additional ancilla qubits for syndrome extraction that could be lost. 

We emphasize that the decoder is applicable regardless of the particular noise model. With different noise models, one will, however, have to reevaluate the application and performance of the decoder. For circuit-based implementations, for instance, data qubits are not measured during the computation and heralded losses of data qubits thus only have a clear physical meaning for the readout of a logical qubit. Furthermore, decoding also can be achieved in the presence of circuit-level noise by repeated syndrome measurements and adding a temporal dimension to the syndrome/Tanner graph~\cite{Dennis2002, Brown2023}. The decoder can then be applied to this higher-dimensional graph in the same way.

\subsection{Decoding topological codes}
In this section, we describe our modifications to the union-find decoder for topological codes. Assume that there are several non-zero syndromes $\sigma_i\in\mathcal{P}_X$ caused by phase-flip errors as illustrated in Fig.~\ref{fig_intro}(a). To perform syndrome validation, we initialize the following data structure: a list $L$ to be used in the breadth-first graph traversal is initialized as $L=[\sigma_1,...,\sigma_{n_s}]$ as illustrated in Fig.~\ref{fig_intro}(b). To avoid traversing elements twice, we mark all the nodes $\sigma_1,...,\sigma_{n_s}$ as visited by setting the corresponding elements of an array $v$ to $v[\sigma_i]=1$, with $v[j]=0$ for all other nodes of the Tanner graph. Furthermore, we count the number of invalid clusters by a variable $n_{iv}$ that is initially set to the number of non-zero syndromes $n_s$ as all these syndromes are isolated clusters in the beginning. As we describe below, the decoder works by starting breadth-first Tanner graph traversal from non-zero syndromes, gradually setting $v[j]=1$, and merging clusters when they meet. This is done until all syndromes are part of a valid cluster ($n_{iv}=0$). Upon completion of this syndrome validation, the syndromes $\sigma_i$ can be explained by errors on only qubits $j$ for which $v_j=1$. Only these qubits will be passed to the peeling-forest decoder~\cite{Delfosse2020} post the syndrome-validation~\cite{Delfosse2021}.

The union-find decoder exploits the fact that erasures are heralded and so appear at known positions. This is achieved by starting Tanner graph traversal resp. cluster growth at the erased qubits~\cite{Delfosse2020}. Why this is advantageous becomes clear in the case where there are only erasures and no Pauli errors. In this case, all non-zero syndromes have at least one neighbor in the Tanner graph which is an erased qubit. Growing from the erasures $\mathcal{E}=[e_1...e_{n_e}]$ to all distance-one neighbors therefore directly makes all clusters valid and the clusters thus stay local below the bond-percolation threshold of the syndrome graph~\cite{Stace2009, Barrett2010, Delfosse2021}. In our framework, starting the cluster growth from the erasures essentially means initializing the list $L$ to the erased qubits $\mathcal{E}=[e_1...e_{n_e}]$ followed by the syndromes (i.e we set $L=[e_1...e_{n_e}\sigma_1...\sigma_{n_s}]$, $\forall_{l\in L}v[l]=1$).

To keep track of connected clusters, we use the union-find tree data structure from Refs.~\cite{Galler1964, Tarjan1975} with a particular implementation that has been developed in the context of percolation theory~\cite{NewmanZiff2001}. Every cluster has a root node where its size and its parity (even or odd number of non-zero syndromes) are stored. Initially, every node is its own root node. When two clusters are merged, the root node $r_1$ of the smaller one is attached to the root node $r_2$ of the larger one (weighted union) by the operation $merge(r_1, r_2)$. The operation $merge()$ also updates the size and the parity of the merged cluster as well as the number of invalid clusters $n_{iv}$ which is decreased by two when two invalid clusters merge and form a valid cluster. The root node $r$ of a cluster can be determined efficiently by an operation $find(r)$ that recursively goes upwards in the tree data structure until the root node is found. More information about the used data structure can be found in Refs.~\cite{NewmanZiff2001, Lobl2023_alg}.

Having the initial data structure in place, the most simple syndrome-validation algorithm works as follows: we loop through the nodes of $L$ with a running variable $i$ specifying the current position in $L$ with $L[i]..L[\text{end}]$ being the active part of $L$. For every node $L[i]$ that is parsed, we loop through all its neighbors. If a neighbor $n$ is not yet in $L$ ($v[n]=0$), we append it to $L$ and mark it as visited by setting $v[n]=1$. If $L[i]$ and its neighbor $n$ belong to different clusters ($root(L[i])\neq root(n)$), we merge the two clusters with $merge()$. The loop terminates once there are no invalid clusters anymore ($n_{iv}=0$). This algorithm is a combination of breadth-first graph traversal of the Tanner graph and updating/merging clusters using the union-find data structure.

\begin{algorithm}[!t]
\caption{Simple union-find decoder for topological codes}\label{alg:1}
\KwIn{$n_e$ erasures $[e_1...e_{n_e}]$ and $n_s$ non-zero syndromes $[\sigma_1...\sigma_{n_s}]$}
\tcc{initialize data structure}
$L=[e_1...e_{n_e}\sigma_1...\sigma_{n_s}]$\\
i = 0\tcp*[l]{position in the list $L$}
\For{$l$ in $L$}{$v[l]=1$}
\tcc{Tanner graph traversal}
\While{there is an invalid cluster}{
    \For{all neighbors $n$ of $L[i]$}{
        \If{$root(L[i]) \neq root(n)$}{
            $merge(root(L[i]), root(n))$
        }
        \If{$v[n]=0$}{
            add $n$ to $L$\\
            $v[n]=1$
        }
    }
    i = i + 1
}
\end{algorithm}

The pseudo-code for the described procedure is given by Algorithm~\ref{alg:1}. This algorithm is similar to an algorithm that has been suggested for decoding general qLDPC codes that can be non-topological codes~\cite{Delfosse2022}. However, we find in section~\ref{sec_results} that the performance of such a simplified decoding algorithm is much worse in terms of logical error rate than the original version of the union-find decoder~\cite{Delfosse2021}. The reason is that also valid clusters are grown in a breadth-first way until all clusters are valid (see Fig.\ref{fig_intro}(b)). Furthermore, Algorithm~\ref{alg:1} may not give rise to an actual threshold for topological codes as we discuss in Appendix~\ref{sec_issue_alg1}. Loosely speaking, the issue is that there is a non-zero probability of having error patterns somewhere in the code for which many steps on the Tanner graph are required in syndrome validation and this probability increases with increasing code size. Problematically, Algorithm~\ref{alg:1} grows all other clusters (also the valid ones) by this number of steps, which can lead to large percolating clusters and associated logical errors. 

\begin{figure*}[!t]
\includegraphics[width=1.0\textwidth]{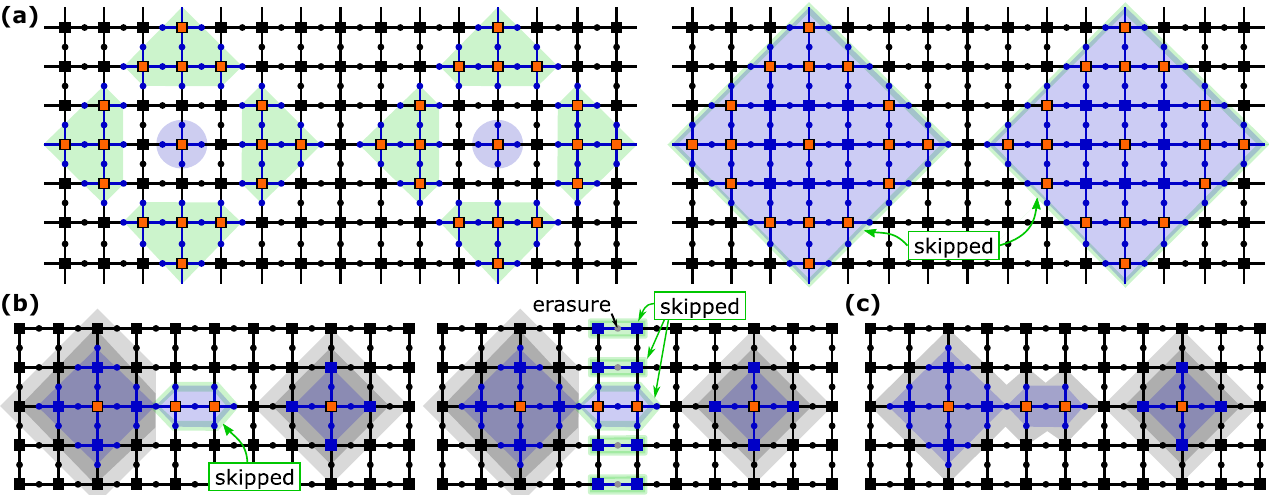}
\caption{\label{fig_issues} Issues associated with skipping nodes in the cluster growth. \textbf{(a)} Possible issue for the method from Fig.~\ref{fig_intro}(c). Two clusters containing an unpaired non-zero syndrome each (blue) grow breadth-first and merge with the surrounding valid clusters (left, green). Eventually, two large invalid clusters are formed (right, blue). Unless skipped nodes are recovered, these clusters will never merge as they are encircled by nodes that have been skipped before (right, green). \textbf{(b)} Left: two invalid clusters are separated by a valid one. One invalid and the valid cluster merge forming a larger non-valid cluster. The nodes of the valid cluster were skipped before and if they are not recovered to the cluster growth (marked with different gray tones) this leads to a larger overall cluster. Right: erased qubits lead to many small valid clusters that are skipped in the breadth-first graph traversal. Unlike the case in (a), this forms a barrier that is non-local and leads to cluster growth in unfavorable directions and eventually logical errors. \textbf{(c)} Algorithm~\ref{alg:3} solves the issues described in (b). Skipped nodes that were part of the previous valid cluster are recovered to the list $L$ upon merging clusters and subsequently participate in the cluster growth.}
\end{figure*}

A simple strategy to circumvent the problems of Algorithm~\ref{alg:1} is skipping a node in the list $L$ if it belongs to a valid cluster. This method avoids the problem of growing valid clusters further as illustrated in Fig.~\ref{fig_intro}(c). An issue with skipping nodes, however, is that this method fails to make all clusters valid in some cases. A corresponding boundary case is illustrated in Fig.~\ref{fig_issues}(a) where several valid clusters encircle two unpaired non-zero syndromes. Growing the invalid clusters breadth-first, one eventually obtains two larger invalid clusters for which none of its boundary vertices is in the active part of $L$ (since these vertices have been skipped before). As a result, these clusters will never become valid. Some modification to recover previously skipped nodes is thus required to avoid this issue.

One option would be storing every skipped node in a second list $L'$. In the rare case of Fig.~\ref{fig_issues}(a), one arrives at the end of $L$ yet there are remaining unpaired clusters. In this case, one sets $L=L'$ and continues breadth-first graph traversal at the beginning of $L$. This method would, however, have obvious disadvantages: assume two far-apart invalid clusters with a valid cluster in between. The boundary vertices of the originally valid cluster will not participate in the cluster growth and syndrome validation will take more steps until the two invalid clusters merge. Furthermore, many small valid clusters with a skipped boundary can even form a partial barrier (not a complete barrier like in Fig.~\ref{fig_issues}(a)) leading to cluster growth in unfavorable directions. These situations are illustrated in Fig.~\ref{fig_issues}(b). We present a numerical investigation of this issue in Appendix~\ref{apdx_b}.

To avoid such issues, recovery of skipped nodes can be obtained by a different algorithm: we store every skipped node $v$ in a cluster-specific list $L'_{root(v)}$ that the root node of the corresponding cluster points to. Once the valid cluster with root $root(v)$ merges with another invalid cluster, the nodes in $L'_{root(v)}$ are restored by attaching them to $L$. The corresponding pseudo-code is given by Algorithm~\ref{alg:3}. This algorithm has the advantage that skipped nodes directly participate in the cluster growth again once they belong to an invalid cluster. This advantage can be seen by comparing Figs.~\ref{fig_issues}(b, c). For Algorithm~\ref{alg:3} (Fig.~\ref{fig_issues}(c)), the boundary vertices of the originally valid cluster participate in the growth and all clusters become valid earlier compared to the case shown in the left part of Fig.~\ref{fig_issues}(b). Using Algorithm~\ref{alg:3}, the size of the final valid clusters will, therefore, be smaller, which increases the distance between clusters. This implies a smaller logical error probability as this probability decays about exponentially with the length of the minimum-weight path that connects small clusters to a larger percolating structure~\cite{Meister2024}.

\begin{algorithm}[!t]
\caption{Improved union-find decoder for topological codes}\label{alg:3}
\KwIn{$n_e$ erasures $[e_1...e_{n_e}]$ and $n_s$ non-zero syndromes $[\sigma_1...\sigma_{n_s}]$}
\tcc{initialize data structure}
$L=[e_1...e_{n_e}\sigma_1...\sigma_{n_s}]$\\
\For{$l$ in $L$}{$v[l]=1$}
\For{$n$ in Tanner graph nodes}{$L'_n=\emptyset$ \tcp*[l]{lists of skipped nodes}}
i = 0\tcp*[l]{position in the list $L$}
\tcc{One Tanner graph step from erasures}
\While{$i < n_e$}{
    \For{neighbors $n$ of $L[i]$}{
        \If{$root(L[i]) \neq root(n)$}{
            $merge(root(L[i]), root(n))$
        } \If{$v[n]=0$}{
                add $n$ to $L$\\
                $v[n]=1$
        }
    }
    i = i + 1
}
\tcc{Further Tanner graph traversal}
\While{there is an invalid cluster}{
    \uIf{$root(L[i])$ is invalid}{
        \For{neighbors $n$ of $L[i]$}{
            \If{$root(L[i]) \neq root(n)$}{
                \If{$L'_{root(n)}\neq\emptyset$}{
                    append $L'_{root(n)}$ to $L$\\
                    $L'_{root(n)}=\emptyset$
                }
                $merge(root(L[i]), root(n))$
            }\If{$v[n]=0$}{
                add $n$ to $L$\\
                $v[n]=1$
            }
        }
    }
    \Else{
        append $L[i]$ to $L'_{root(L[i])}$
    }
    i = i + 1
}
\end{algorithm}

\subsection{Decoding general qLDPC codes}
For non-topological quantum low-density-parity-check (qLDCP) codes, the described algorithms do not work since a qubit can be part of more than two parity checks~\cite{Bravyi2024}. Therefore, deciding whether a cluster is valid/invalid via its parity is not possible. There exist algorithms for decoding such codes~\cite{Panteleev2021b, Roffe2020, Iolius2024, Wolanski2024} but they typically cannot handle erasures. The decoder from Ref.~\cite{Delfosse2022} can, but we expect that this algorithm has issues similar to our Algorithm~\ref{alg:1} for topological codes (see Appendix~\ref{sec_issue_alg1}). We, therefore, give a modification of Algorithm~\ref{alg:3} that can be used for decoding arbitrary CSS codes in the presence of both Pauli errors and erasures.

First, we need to change the method used for syndrome validations of clusters. To this end, we solve the equation system $\sigma_{cl}=H_{cl}\cdot e_{cl}$ by Gaussian elimination when new nodes are added to the cluster. Here, $\sigma_{cl}, e_{cl}$ refer to all syndromes and qubits that are part of the cluster. $H_{cl}$ is the parity-check matrix shrunk to only the nodes of this cluster. If a solution is found, the cluster is valid and the solution already is a final decoding unless being overwritten by a new solution should the cluster merge with another one. If no solution can be found, the cluster is invalid.

Second, we need to ensure that a solution $e_{cl}$ does not flip any syndromes that are not part of the cluster. For the standard union-find decoder, this is ensured by decoding clusters with the peeling forest method~\cite{Delfosse2020}. Here we ensure this by growing all clusters such that their boundary consists of only syndromes $\mathcal{P}_X$ or $\mathcal{P}_Z$ and no data qubits $\mathcal{Q}$. After initially growing from all erased qubits to all their neighbors by one step on the Tanner graph, this is automatically fulfilled. For the subsequent growth, we make sure that this condition is satisfied by adding all first- and second-order neighbors of a syndrome to a cluster in one step: add all neighbors of the syndrome (which are data qubits) and then add all neighbors of these neighbors (which are syndromes again). The corresponding pseudo-code is given as Algorithm~\ref{alg:4}.

\begin{algorithm}[!t]
\caption{Decoder for qLDPC codes}\label{alg:4}
\KwIn{$n_e$ erasures $[e_1...e_{n_e}]$ and $n_s$ non-zero syndromes $[\sigma_1...\sigma_{n_s}]$}
\tcc{initialize data structure}
$L=[e_1...e_{n_e}\sigma_1...\sigma_{n_s}]$\\
\For{$l$ in $L$}{$v[l]=1$}
\For{$n$ in Tanner graph nodes}{$L'_n=\emptyset$ \tcp*[l]{lists of skipped nodes}}
i = 0\tcp*[l]{position in the list $L$}
\tcc{One Tanner graph step from erasures}
\While{$i < n_e$}{
    \For{neighbors $n$ of $L[i]$}{
        \If{$root(L[i]) \neq root(n)$}{
            $merge(root(L[i]), root(n))$
        } \If{$v[n]=0$}{
                add $n$ to $L$\\
                $v[n]=1$
        }
    }
    $validate(root(L[i]))$\\
    i = i + 1
}
\tcc{Further Tanner graph traversal}
\While{there is an invalid cluster}{
    \uIf{$root(L[i])$ is invalid}{
        \For{all neighbors $n$ of $L[i]$}{
        \If{$root(L[i]) \neq root(n)$}{
            $merge(root(L[i]), root(n))$ \\
            \For{all neighbors $n2$ of $n$}{
                \If{$L'_{root(n2)}\neq\emptyset$}{
                    append $L'_{root(n2)}$ to $L$\\
                    $L'_{root(n2)}=\emptyset$
                }
                \If{$root(L[i]) \neq root(n2)$}{
                    $merge(root(L[i]), root(n2))$
                }
                \If{$v[n2]=0$}{
                    add $n2$ to $L$\\
                    $v[n2]=1$
                }
            }
        }
        }
        $validate(root(L[i]))$
    }
    \Else{
        append $L[i]$ to $L'_{root(L[i])}$
    }
    i = i + 1
}
\end{algorithm}

In Algorithm~\ref{alg:4}, the procedure $merge()$ merges two clusters, and updating the validity of the merged cluster is done with a separate function $validate()$ that solves the equation system $\sigma_{cl}=H_{cl}\cdot e_{cl}$ via Gaussian elimination. The function $validate()$ takes the root node index of the cluster as an argument and validates/decodes the cluster. Updating the number of invalid clusters $n_{iv}$ is done in the following way: $merge(root(L[i]), root(n))$ decrements $n_{iv}$ by one if $n$ belongs to an invalid cluster as this independent invalid cluster disappears upon the merge. $validate(root(L[i]))$ decrements $n_{iv}$ by one if the originally invalid cluster becomes valid by merging with other clusters.

The advantage of separating $validate()$ and $merge()$ is that the time-consuming Gaussian elimination is not performed for every merging event, but only after having merged all first- and second-order neighbors of the current node in the list $L$. For Algorithm~\ref{alg:4}, a remaining issue is that Gaussian elimination with time complexity $O(N^3)$ is performed multiple times, leading to an upper bound for the time complexity of $O(N\cdot N^3)$ (assuming that Gaussian elimination is performed $\sim N$ times for clusters of size $\sim N$). Below the error threshold, however, the average size of the clusters is much smaller than the system size~\cite{Griffiths2023}. For only erasures, the size of the largest connected component would for instance scale as $log(N)$ according to a result from percolation theory~\cite{Duminil2018}. This bounds the complexity spent on a single cluster by multiple Gaussian eliminations to $O(log(N)^4)$. As there can be at most $N/log(N)$ such clusters, the typical time complexity will be below $O(N/log(N)\cdot log(N)^4)=O(N\cdot log(N)^3)$. Numerically, we find in the next section that the scaling of Algorithm~\ref{alg:4} stays close to linear in the practically most relevant sub-threshold regime.

\section{Results}
\label{sec_results}
\subsection{Error and erasure thresholds for topological codes}
\begin{figure}[!t]
\includegraphics[width=\linewidth]{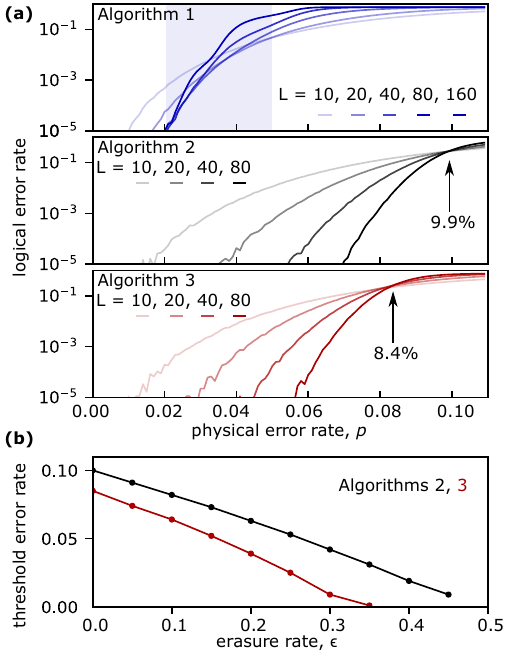}
\caption{\label{fig_resuts} \textbf{(a)} Logical error rate of the two-dimensional surface code as a function of the Pauli error rate $p$ assuming zero erasures ($\epsilon=0$). For Algorithm~\ref{alg:1}, curves of different code sizes cross in the regime shaded in blue yet no clear threshold can be identified. For the two other decoding algorithms, the threshold error rate can be determined by the crossing point of the curves for different code sizes (see arrows). Below these crossing points, the logical error rate decreases as we increase the code size. Algorithm~\ref{alg:4} has worse performance than Algorithm~\ref{alg:3} but is still interesting since it can also be applied to non-topological qLDPC codes. \textbf{(b)} Threshold error rate as a function of the erasure rate $\epsilon$. Data points correspond to numerical simulations and the lines are guides to the eye.}
\end{figure}

We benchmark our algorithms on the two- and (2+1)-dimensional surface codes with periodic boundary conditions (toric codes). For the two-dimensional surface code, we assume perfect syndrome measurements (see section~\ref{sec_noise}). The (2+1)-dimensional surface code corresponds to a syndrome graph with an additional time direction ($+1$) where syndrome measurements are repeated and we assume the same error rate $p$ for Pauli errors on data qubits and errors in the syndrome measurements. This model is equivalent to measurement errors in the Raussendorf-Harrington-Goyal lattice~\cite{Raussendorf2006} for measurement-based quantum computing.

For zero erasure rate, the dependence of logical versus physical error rate of the two-dimensional surface code is shown in Fig.~\ref{fig_resuts}(a) for the three algorithms that we have developed. Algorithm~\ref{alg:3} shows the best performance in terms of logical error rate. Algorithm~\ref{alg:4} performs slightly worse as Tanner graph traversal is done in double steps to ensure that the cluster boundaries consist of only syndromes and no data qubits\footnote{For the simulations where Algorithm~\ref{alg:4} is applied to simulate thresholds of surface codes (Fig.~\ref{fig_resuts}(a, b)), we have exploited that a valid cluster merged with an invalid one gives a larger invalid cluster. This speeds up the simulation without changing the thresholds compared with the full version of Algorithm~\ref{alg:4}.}. For both of these algorithms, we find threshold error rates below which the logical error rate decreases as we make the code size larger. This indicates that we can make the logical error rate arbitrarily low by choosing a sufficiently large code size. In comparison, Algorithm~\ref{alg:1} performs significantly worse and shows no clear threshold. This is because the algorithm also grows valid clusters as described above.

Under the assumptions used here, the error correction thresholds as obtained by a minimum perfect weight matching (MWPM) decoder are $0.103$ and $0.029$, for the two- and (2+1)-dimensional surface code, respectively~\cite{Dennis2002, Wang2003}. When using Algorithm~\ref{alg:3}, we find a threshold of $0.099$ for the two- and a threshold of $0.026$ for the (2+1)-dimensional surface code. The threshold values are identical to the values obtained by the original union-find decoder~\cite{Delfosse2021} despite the simplifications that our Algorithm~\ref{alg:3} makes. The agreement of the thresholds is remarkable, as the algorithm from Ref.~\cite{Delfosse2021} differs conceptionally from ours. The decoder from Ref.~\cite{Delfosse2021} grows cluster-by-cluster and sorts the clusters by the size of their boundary, which boosts the threshold. We avoid tracking cluster boundaries and grow node-by-node instead.

Finally, we investigate the error correction threshold as a function of the erasure rate $\epsilon$. For the two-dimensional surface code, these simulations are shown in Fig.~\ref{fig_resuts}(b) for the different algorithms, except Algorithm~\ref{alg:1} for which no clear threshold can be obtained. In the absence of erasures, these simulations correspond to the Pauli error thresholds determined in Fig.~\ref{fig_resuts}(a). In the absence of Pauli errors, erasures up to the bond-percolation threshold of the syndrome graph can be tolerated. Fig.~\ref{fig_resuts}(b) shows that this is the case for Algorithm~\ref{alg:3}. For Algorithm~\ref{alg:4}, this is also the case in the complete absence of Pauli errors ($p=0$). However, in the regime above $\epsilon\sim 0.35$, even an extremely low Pauli error rate $p$ cannot be tolerated. The reason for this issue is the following: after the initial part of the algorithm where one step on the Tanner graph starting from the erasures is performed, Algorithm~\ref{alg:4} performs further Tanner graph traversal in double steps to validate the remaining clusters that are invalid due to Pauli errors. This can lead to a large percolating cluster already after making one double step for every node in the active part of $L$. For Algorithm~\ref{alg:3}, this issue does not appear as cluster growth is performed by traversing the Tanner graph in single steps. If the error probability is sufficiently low, all invalid clusters will become valid before a percolating cluster is created even if the erasure rate is high. Overall, the results show that the developed algorithms are capable of efficiently handling a combination of loss and logical errors, with Algorithm~\ref{alg:3} having a slightly better accuracy. 

\subsection{Running times for topological codes}
\begin{figure}[!t]
\includegraphics[width=\linewidth]{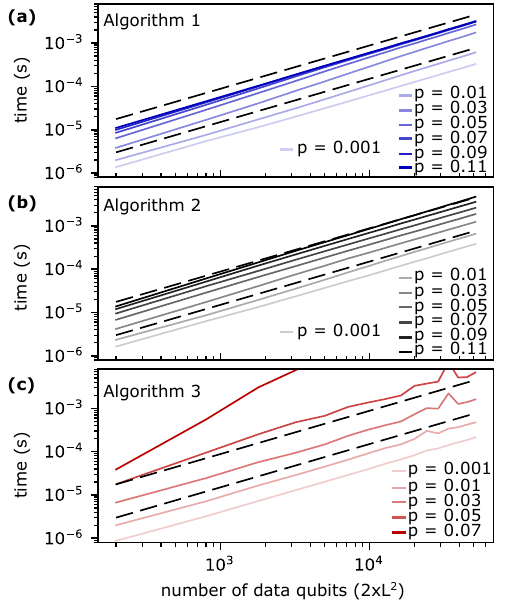}
\caption{\label{fig_time} Decoding time as a function of code size (used processor: Intel(R) Core(TM) i7-10700K CPU @ 3.80GHz, used system: Ubuntu 22.04.4 LTS). For various Pauli error rates $p$, the running time [averaged over $3\cdot10^3$ repetitions] scales linearly with the code size as specified by the number of data qubits $2\times L^2$). The dashed black curves indicate a linear reference and are identical for all sub-plots. Note that some Pauli error rates are above the corresponding error threshold.}
\end{figure}
Compared to other decoding algorithms like minimum-weight matching~\cite{Higgott2023}, the original union-find decoder from Ref.~\cite{Delfosse2021} has the advantage that the running time scales linearly with the code size\footnote{Ref.~\cite{Delfosse2021} has shown that their algorithm has a running time of $O(N\alpha(N))$ where the inverse Ackermann's function $\alpha(N)$ does not increase above $3$ for a system of any realistic size.}. Algorithm~\ref{alg:1} preserves this linear scaling as every node of the Tanner graph is only visited once in the breadth-first graph traversal. For Algorithm~\ref{alg:3} a linear scaling is not directly obvious since vertices can be skipped and later added again to the list $L$. That this process happens multiple times for many nodes is, however, unlikely since the skipped nodes are added to the end of $L$ and so all invalid clusters are grown before these nodes are reconsidered. We find numerically for the surface code that the overall number of nodes encountered in the breadth-first graph traversal is always below twice the number of nodes in the Tanner graph for more than $2\cdot 10^7$ samples of a broad parameter regime\footnote{As there are never more than $N$ active elements in $L$ at the same time, the length of $L$ can at most be $N$ and we start over at position zero when the end of the list is reached.}.

For the discussed reasons, we expect that the running times of Algorithm~\ref{alg:1} and Algorithm~\ref{alg:3} scale linearly with the system size $N$, or the running time is at least asymptotically bounded by a linear curve. This expectation is numerically confirmed for various rates of Pauli errors on the two-dimensional surface code as shown in Figs.~\ref{fig_time}(a,b). Algorithm~\ref{alg:3} achieves the same threshold as the decoder from Ref.~\cite{Delfosse2020} and decodes the two-dimensional surface code with code distance $10$ in $~\sim 2\mu s$ on a general-purpose computer (for a Pauli error rate of $0.01$). This is faster by about a factor of five compared to the benchmark presented in Ref.~\cite{Delfosse2020}. As no optimizations of our implementations on a technical level have been performed, the shorter running time may be caused by the simplification of our algorithms that avoid tracking cluster boundaries, but the differences in implementation and hardware make it hard to draw any firm conclusions.

Finally, we investigate the running time of Algorithm~\ref{alg:4} when applied to the surface code. As the surface code can be more efficiently decoded by e.g. Algorithm~\ref{alg:3}, the purpose of this analysis is merely to compare the speed of Algorithm~\ref{alg:4} to the other algorithms. The running time as a function of system size is plotted in Fig.~\ref{fig_time}(d) for different error rates. We find that a linear scaling with the system size is maintained at error rates that are far below the threshold. Once the error rate gets close to the threshold, this linear scaling breaks down due to the non-linear time complexity of using Gaussian elimination for decoding large clusters.

\subsection{Error thresholds and running time for non-topological qLDPC codes}
\begin{figure}[!t]
\includegraphics[width=\linewidth]{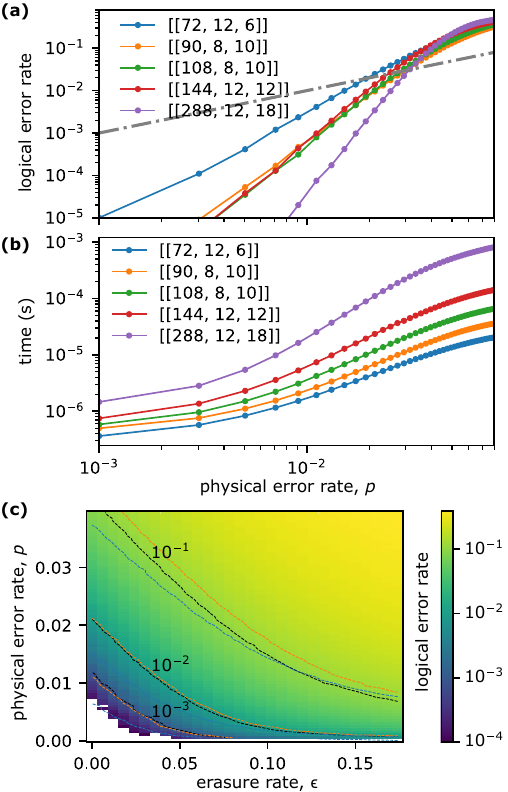}
\caption{\label{fig_ldpc} Performance of Algorithm~\ref{alg:4} on non-topological qLDPC codes assuming perfect syndrome measurements. \textbf{(a)} Logical vs. physical error rate for several bivariate bicycle codes from Ref.~\cite{Bravyi2024}. The codes are represented by their parameters $[[n, k, d]]$ where $n$ is the number of data qubits, $k$ is the number of logical qubits, and $d$ is the code distance. The dashed gray line indicates where logical and physical error rates coincide. Data points are an average of $10^6$ repetitions. \textbf{(b)} Average running time of a single decoding for the Pauli error rates from (a). Used processor: Intel(R) Core(TM) i7-10700K CPU @ 3.80GHz, used system: Ubuntu 22.04.4 LTS. \textbf{(c)} Logical error rate as a function of the erasure rate and the physical Pauli error rate. The colors correspond to the logical error rate of the $108$-qubit code from (a) and the black contour lines mark a constant logical error rate of the indicated value. The blue and orange lines are corresponding contour lines for the $72$- and $90$-qubit codes. The data points are averaged over $10^5$ samples and we cut the data at a logical error rate of $10^{-4}$ for better visibility.}
\end{figure}
Topological codes like the surface code have low-density parity checks which can be implemented via local interactions making them well-suited for many hardware implementations. However, they have a poor encoding rate meaning that the number of encoded logical qubits $k$ is very small compared to the number of physical qubits $n$. For the surface code, for instance, the encoding rate $k/n$ is asymptotically zero for increasing $n$ as the number of logical qubits $k$ is constant. Furthermore, the code distance $d$ only scales sub-linearly with the system size as $d\sim\sqrt{n}$. These issues have motivated the development of new qLDPC codes with an asymptotically constant rate~\cite{Panteleev2022, Bravyi2024} and better scaling of the code distance~\cite{Panteleev2022}.

In this section, we use Algorithm~\ref{alg:4} to decode such non-topological qLDPC codes. In particular, we consider a class of bivariate bicycle codes from Ref.~\cite{Bravyi2024}. Fig.~\ref{fig_ldpc}(a) shows simulations of logical versus physical error rates assuming perfect syndrome measurements. From the shown simulations, we obtain so-called pseudo-thresholds as the point where the logical error rate (probability that at least one of the logical qubits has an error) coincides with the physical error rate\footnote{This is a conservative estimate and for $k$ encoded logical qubits, one could instead look for the error rate $p$ for which the logical error rate coincides with the probability $1-(1-p)^k\sim p\cdot k$ that at least one of $k$ physical qubits has an error~\cite{Bravyi2024}. This would yield a higher pseudo-threshold.}. For the $72$-, $90$-, $108$-, $144$-, $288$-qubit codes from Ref.\cite{Bravyi2024}, we find for the corresponding pseudo-thresholds $0.019, 0.030, 0.028, 0.025, 0.031$ (where we have taken the worst value from the simulations involving $H_X$ and $H_Z$). For these simulations, Fig.~\ref{fig_ldpc}(b) displays the average running time of a single decoding. For a Pauli error rate of $p=0.001$, we find a sub $1\,\mu s$ decoding time for most of the considered codes. This decoding speed is already faster than the time for a single syndrome extraction cycle in state-of-the-art superconducting qubit devices~\cite{Krinner2022, Google2023}. At larger error rates, the decoder slows down more strongly than for the previously considered topological codes. Again, this is due to the use of Gaussian elimination for decoding: higher error rates lead to larger cluster sizes, which have a stronger impact on the running time when Gaussian elimination is used.

Finally, we apply Algorithm~\ref{alg:4} to the qLDPC codes in the presence of Pauli errors as well as erasures. The capability of including qubit erasure in the decoding process is relevant for all platforms where qubits can be lost. This applies to photonic quantum computing architectures~\cite{Bartolucci2021, Paesani2022} as well as architectures based on Rydberg atoms~\cite{Bluvstein2024} or trapped ions~\cite{Stricker2020} -- all platforms compatible with long-distance connectivity and thus well-suited for implementing qLDPC codes. Again, we consider the codes from Ref.~\cite{Bravyi2024} for which Fig.~\ref{fig_ldpc}(c) shows a simulation of the logical error rate as a function of qubit erasure and Pauli errors. As seen in the figure, the decoder performs as desired and is capable of correcting a combination of Pauli and erasure errors for the considered non-topological qLDPC codes.

\section{Summary and Outlook}
In summary, we have suggested and implemented simplifications of the original union-find decoder~\cite{Delfosse2021} for topological codes. We have demonstrated a fast running time of the decoder due to a simplification of the original algorithm which is a potential advantage for fast real-time decoding. Importantly, our results indicate that error correction thresholds are not affected by the simplifications. Furthermore, we have developed a union-find algorithm suitable for decoding non-topological qLDPC codes. This decoder can be applied to arbitrary CSS codes in the presence of both Pauli errors and erasures.

For future work, it would be interesting to investigate to which extent further simplifications or improvements can be made. Removing weighted union and path compression~\cite{NewmanZiff2001} in the operation $merge()$ has, for instance, been investigated~\cite{Griffiths2023, Liyanage2023}. To account for non-uniform error probabilities, it would be interesting to investigate how weighted edges~\cite{Huang2020, Huang2020b, Liyanage2024} can be included in our implementation. For our union-find decoder for qLDPC codes, it would be desirable if Gaussian elimination could be avoided or if the number of Gaussian eliminations could be reduced by updating the validity of clusters less frequently. Alternatively, one may employ methods such as the on-the-fly elimination recently proposed in Ref.~\cite{Hillmann2024}. For fast real-time decoding, parallelization is eventually relevant and data structures are ideally kept small and local~\cite{Liyanage2023, Chan2023} to enable parallelization and integration on FPGA or ASIC hardware~\cite{Das2022, Liyanage2023, Barber2023, Liyanage2024}. This is in slight conflict with the fact that our simplification relies on using a global data structure for breadth-first graph traversal. However, this issue may be circumvented by embedding our decoder into an architecture with local decoding windows decoded by an inner decoder and boundaries being fixed by an outer decoder~\cite{Skoric2023, Tan2023}. Complementary approaches could be using local pre-decoding strategies~\cite{Delfosse2020b, Alavisamani2024, Smith2023} or using our decoder for small inner codes that are concatenated with outer codes~\cite{Meister2024}. For the latter approach, a soft output identifying cases with high logical error probability has turned out to be very useful~\cite{Meister2024, Smith2024}. Soft output or postselecting low-Hamming-weight syndromes~\cite{Vittal2023} may also help circumvent the discussed issues of Algorithm~\ref{alg:1} and could be added relatively easily to our implementation.

\section{Acknowledgements}
We would like to thank Love A. Pettersson, Nicolas Delfosse, and Aliki Capatos for useful discussions. We are grateful for financial support from Danmarks Grundforskningsfond (DNRF 139, Hy-Q Center for Hybrid Quantum Networks) and Innovation Fund Denmark (IFD1003402609, FTQP). S.P. acknowledges funding from the Marie Skłodowska-Curie Fellowship project QSun (nr. 101063763), from the VILLUM FONDEN research grant No.VIL50326 and No.VIL60743, and support from the NNF Quantum Computing Programme. S.X.C. acknowledges support from UK EPSRC (EP/SO23607/1).

\appendix
\renewcommand\thefigure{\thesection.\arabic{figure}}  

\section{Appendix: Issues with Algorithm 1}
\label{sec_issue_alg1}
In this section, we explain the issue that prevents Algorithm~\ref{alg:1} from having a threshold such that below the threshold an arbitrarily low logical error rate could be obtained by making the code size larger and larger. The problem is the following: assume a given error rate $p>0$. For any value of $p$, we can find a size of a code patch $s_p$ such that the probability of finding at least one non-zero syndrome inside the patch is larger than the bond-percolation threshold $\lambda_{bond}$ of the syndrome graph\footnote{For the two-dimensional surface code, the syndrome graph is a two-dimensional square lattice such that $\lambda_{bond}=1/2$.}. Once a corresponding patch size $s_p$ is chosen, we can always find a finite number of patches $N_p$ such that with a high probability, there is at least one patch with an error that spans from one end of the patch to the other, creating two far apart non-zeros syndromes while all surrounding patches do not have a non-zero syndrome. Since this is a structure of finite size, such a patch occurs with a certain probability corresponding to a fixed density, independent of the code size. To correct such an error, Algorithm~\ref{alg:1} needs to grow a cluster of the size of the entire patch. Such a cluster connects the patch to all its neighboring patches. The problematic thing is that clusters (also valid clusters) are grown by the same size. This applies to all syndrome-containing patches and by our assumption about $s_p$ their fraction among all patches is larger than $\lambda_c$. Since all these clusters leak into the corresponding neighboring patches after the growth, a super-cluster spanning the entire lattice will be grown, which can lead to a logical error. Given a small value for $p$, the required patch sizes $s_p$ as well as the number of patches $N_p$ may be very large, potentially leading to an extremely large number of qubits $n\sim N_ps_p$ before this issue emerges. However, the described argument shows that the issue will eventually appear, and there is no threshold $\lambda_e$ in the sense that for all values $p<\lambda_e$, a larger code size can reduce the logical error rates to arbitrarily low values. The logical error rate may decrease up to some code sizes but then increase again for sizes of $n\sim N_ps_p$ or more\footnote{The same argument may apply to a similar algorithm presented in Ref.~\cite{Delfosse2022} but further investigations are needed to fully resolve this. For the algorithm from Ref.~\cite{Delfosse2022}, errors with Hamming weight bounded by $An^{\alpha}$ are shown to be successfully corrected~\cite{Delfosse2022}. This does not contradict our argument as $\alpha<1$ and thus the weight of provably correctable errors vanishes compared to the overall number of qubits $n$.}.

\section{Appendix: variant of Algorithm~\ref{alg:3}}
\label{apdx_b}
Here, we briefly discuss an alternative implementation of Algorithm~\ref{alg:3} where skipped nodes are only recovered to participate in the cluster growth in case the algorithm reaches the end of the list $L$ yet invalid clusters are still present. The corresponding pseudo-code is given as Algorithm~\ref{alg:2b}.

\begin{algorithm}[!t]
\caption{variant of Algorithm~\ref{alg:3}}\label{alg:2b}
\KwIn{$n_e$ erasures $[e_1...e_{n_e}]$ and $n_s$ non-zero syndromes $[\sigma_1...\sigma_{n_s}]$}
\tcc{initialize data structure}
$L=[e_1...e_{n_e}\sigma_1...\sigma_{n_s}]$\\
i = 0\tcp*[l]{position in the list $L$}
\For{$l$ in $L$}{$v[l]=1$}
$L' = [\,]$\tcp*[l]{list of skipped nodes}
\tcc{Tanner graph traversal}
\While{there is an invalid cluster}{
    \If{$i=len(L)$}{
        $L=L'$\\
        $i=0$
    }
    \uIf{$root(L[i])$ is invalid \textbf{or} $L[i]\in\mathcal{E}$}{
        \For{all neighbors $n$ of $L[i]$}{
            \If{$root(L[i]) \neq root(n)$}{
                $merge(root(L[i]), root(n))$
            }
            \If{$v[n]=0$}{
                add $n$ to $L$\\
                $v[n]=1$
            }
        }
    }
    \Else{
        append $L[i]$ to $L'$
    }
    i = i + 1
}
\end{algorithm}

\begin{figure}[!t]
\includegraphics[width=\linewidth]{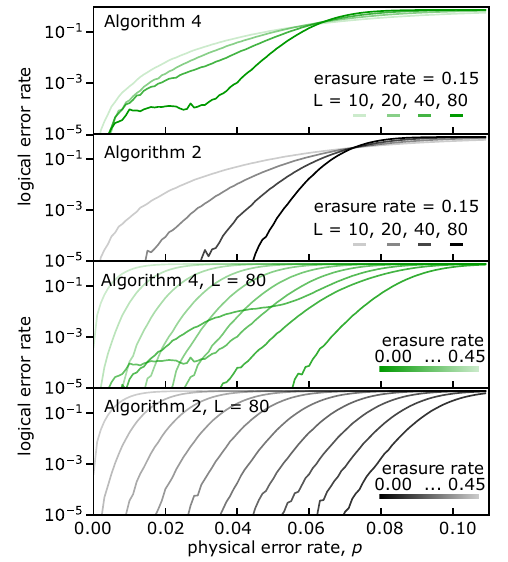}
\caption{\label{fig_suppl} Problems with Algorithm~\ref{alg:2b} at finite erasure rates. Top part: at a fixed erasure rate ($0.15$), a larger code size can increase the logical error probability. Bottom part: at fixed code size ($L=80$), adding more erasures can lead to lower logical error rates. These seemingly contradictory cases can be explained by the issue discussed in Fig.~\ref{fig_issues}(b). The corresponding plots for Algorithm~\ref{alg:3} show that this algorithm suffers none of these problems and is thus our algorithm of choice for decoding topological codes.}
\end{figure}

We find that Algorithm~\ref{alg:2b} gives an error threshold of $0.085$ for the surface code in the regime of zero erasures. However, we find that for erasure rates in the interval $(0,\sim 0.2)$, Algorithm~\ref{alg:2b} does not give a threshold at all. This becomes emergent as effects that are reminiscent of noise floors\footnote{Some decoders are known to suffer from noise floors that set a lower bound for the logical error rate~\cite{Meister2024}. However, the reason is likely different from what we discuss in the context of Algorithm~\ref{alg:2b}.} or effects where an increasing erasure rate leads to fewer logical errors. Numerical results showing these issues are plotted in Fig.~\ref{fig_suppl}. Importantly, Algorithm~\ref{alg:3} and Algorithm~\ref{alg:4} do not have such problems since skipped nodes are recovered to the cluster growth when clusters merge.

The reason for this problem of Algorithm~\ref{alg:2b} is illustrated on the right of Fig.~\ref{fig_issues}(b) where many small valid clusters originating from erased qubits form a barrier of skipped nodes preventing two invalid clusters from merging. Unlike the situation in Fig.~\ref{fig_issues}(a), the growth of invalid clusters does not halt, yet invalid clusters will keep growing in unfavorable directions. This can lead to a cluster that spans the entire lattice, making a logical error likely. Using simulations similar to the ones presented in section~\ref{sec_results}, we find numerically that Algorithm~\ref{alg:2b} gives a threshold for the surface code in the absence of erasures. For a certain range of erasure rates, though, it does not give a threshold below which logical errors can be reduced to arbitrarily low values by increasing the code size. This is due to the issues illustrated in Fig.~\ref{fig_issues}(b) making Algorithm~\ref{alg:2b} only useful in a certain parameter regime. Therefore, Algorithm~\ref{alg:3} is our preferred algorithm for decoding topological codes.

\bibliography{lib}

\end{document}